# Enhanced spin – reorientation temperature and origin of magnetocapacitance in $HoFeO_3$

Ganesh Kotnana[1] and S. Narayana Jammalamadaka[1*]

[1]*Magnetic Materials and Device Physics Laboratory, Department of Physics, Indian Institute of Technology Hyderabad, Hyderabad, India – 502285.*

*Corresponding author: surya@iith.ac.in

**Abstract:**

We report on the increase in the spin reorientation temperature in $HoFe_{0.5}Cr_{0.5}O_3$ compound by isovalent substitution ($Cr^{3+}$) at the Fe-site and the magnetocapacitance in the $HoFeO_3$ compound. Spin reorientation transition is evident around 50 K and 150 K for the $x = 0$ and $x = 0.5$ compounds respectively. The increase in the spin reorientation transition temperature in case of $x = 0.5$ compound can be attributed to the domination of the $Ho^{3+}$- $Fe^{3+}$ interaction over the $Fe^{3+}$- $Fe^{3+}$ interaction. Decrease in Néel temperature from 643 K ($x = 0$) to 273 K ($x = 0.5$) can be ascribed to the decrease in the interaction between antiferromagnetically aligned $Fe^{3+}$ moments as a result of the dilution of the $Fe^{3+}$ moments with the $Cr^{3+}$ addition. From the magnetization M *vs.* magnetic field H variation it is evident that the coercivity, $H_C$ decreases for $x = 0.5$ compound, hinting the magnetic softening of the $HoFeO_3$ compound. Observed magnetocapacitance could be due to lossy dielectric mechanism in the present compound. Indeed, present results would be helpful in understanding the physics behind rare-earth orthoferrites.

**Introduction**:

The complex magnetic coupling between the ions in the materials containing the rare earth and the transition metal shows a variety of interesting properties. Fundamentally, the orientation of the strongly interacting spins of a transition metal which order at high temperatures can be controlled by the spins of the rare earth ions which order at lower temperatures, may lead to the spin reorientation (SR) transitions [1]. Such SR transitions may affect the magnetic, optical and elastic properties of the materials and have been observed in the compounds like rare-earth ferrites, chromites and manganites [2-4].

The rare-earth orthoferrites have the complex spin structure between the rare-earth and the transition metal ions and are distorted perovskite family of canted antiferromagnets drawn considerable attention due to their unique physical properties [5] and potential application such as solid oxide fuel cells [6], magneto-optic devices [7], and multiferroics [8]. The Dzyaloshinsky–Moriya type canted weak ferromagnetism, an unusual large antisymmetric exchange interactions, extremely small anisotropy of Fe spins in $a - c$ plane, and large anisotropy towards $b$ - axis have been are the characteristics of these compounds [9]. Particularly, it has been understood that the $Fe^{3+}$ moments in the $RFeO_3$ compounds undergo a magnetic phase transition into a canted antiferromagnetic state in the temperature interval 620 - 750 K. Among all the characteristics of the $RFeO_3$ compounds, the SR is one of the distinctive magnetic transitions, where the direction of the easy axis of magnetization changes from one crystal axis to the other at lower temperatures [1]. Interestingly, the $HoFeO_3$ compound is a representative of the rare-earth orthoferrites and has been believed to possess a canted G-type antiferromagnetism and potential candidate for the ultrafast recording [10] with a magnetic ordering temperature around 641 K [11]. For example, the $HoFeO_3$ compound exhibits the SR transition around 50 K due to the competition between various complex magnetic interactions. Such a SR transition leads to a change in the magnetic structure from $G_xA_yF_z$ ($\Gamma_4$) to $F_xC_yG_z$ ($\Gamma_2$), where $G_x$, $A_y$, and $F_z$ stand for the spin component along x, y, and z axis respectively in terms of the Bertaut's notation [12]. In recent years, it has been believed that the variety of interesting properties can be induced by alloying different kinds of cations at the B-site of the perovskite materials [13]. According to the Goodenough-Kanamori theory, the $Cr^{3+}$ is best choice for the $Fe^{3+}$ to show superior magnetic properties due to a superexchange interaction [14]. On top of that it has been observed that the multiferroic behaviour in $YCr_{1-x}Fe_xO_3$ [15] and the magnetodielectric effect in the $DyFe_{0.5}Cr_{0.5}O_3$ [16] compounds by having both the Fe and Cr. Such interesting observations have not been paralleled in the Ho based orthoferrites. Substitution of the $Cr^{3+}$ to the $Fe^{3+}$ may give interesting structural and magnetic properties in the $HoFeO_3$ compound. Hence, in the present work, our aim is to study and understand the complex magnetic interactions, tuning the SR transition and study the magnetocapacitance (MC) behaviour of the $HoFeO_3$ compound.

**Experimental Details**:

Polycrystalline compounds of the $HoFe_{1-x}Cr_xO_3$ ($x$ = 0 and 0.5) were prepared by conventional solid state reaction method using the High purity oxide powders of $Ho_2O_3$, $Fe_2O_3$, and $Cr_2O_3$ (purity > 99.9%) (Sigma-Aldrich chemicals India) as starting raw materials. The phase purity or the structural analysis was carried out at room temperature using the powder x - ray diffraction (XRD) (PANalytical X-ray diffractometer) with a Cu-$K_\alpha$ radiation ($\lambda$ = 1.5406 Å) and with a step size of $0.017^0$ in the wide range of the Bragg angles 2θ ($20^0$ - $80^0$). Magnetization (M) *vs.* temperature (T) and M *vs.* magnetic field (H) were recorded with a Quantum Design MPMS-3 in the temperature range of 5 - 300 K. M *vs.* T was measured in the zero field cooling (ZFC) as well as the field cooling (FC) conditions. Essentially, in the ZFC conditions, the compound was cooled from 300 K to 5 K without application of any magnetic field. However, in case of the FC condition, the compound was cooled from 300 K in the presence of 1000 Oe up to a desired temperature, 5 K. Differential scanning calorimeter (DSC) in specific heat mode was used to measure the phase transition above 300 K. During the measurement, the rate of heating was 10 K/min. Room temperature magnetocapacitance (MC) measurements were carried out on the $HoFeO_3$ compound. Here the sample consists 10 mm as diameter and 1.692 mm as thickness. After keeping the $HoFeO_3$ with the above dimensions in between aluminium discs with a diameter of 10 mm, MC was measured using Andeen - Hangeling 2700A Capacitance Bridge at 100 Hz. Room temperature polarization P *vs.* Electric field E loop measurements were carried out using aixACCT TF 2000 analyzer.

**Results and Discussion**:

Phase purity of the compounds $HoFe_{1-x}Cr_xO_3$ ($x$ = 0, 0.5) is confirmed with the powder XRD. From the Fig. 1 it is clear that observed and indexed reflections are allowed for a compound with an orthorhombic $GdFeO_3$ type disordered perovskite structure with a space group *Pbnm*. In addition to parent phase, we also could able to see a small peak which corresponds to iron garnet phase, $Ho_3Fe_5O_{12}$ at around 32°, 35° and 45°. It has been reported that the rare-earth iron garnet ($R_3Fe_5O_{12}$) phase is stable compared to orthoferrites ($RFeO_3$) at high temperature [17, 18]. The refinement has been carried out using the GSAS software and results of which have been published elsewhere [19].

It has been reported that in the rare-earth orthoferrites, each $Fe^{3+}$ ion on an adjacent octahedral site interact though an antiferromagnetic superexchange interaction mediated by an oxygen [14]. The strong interaction between the iron magnetic moments at high temperatures results in the transition from an antiferromagnetic (AFM) state to a paramagnetic state characterized by a Néel temperature ($T_{N2}$). Fig. 2(a) illustrates the temperature dependence of the susceptibility $\chi$ (M *vs.* T) of $HoFe_{1-x}Cr_xO_3$ ($x$ = 0, 0.50) compounds which were measured under ZFC condition in the temperature range 5 – 300 K. For $x$ = 0 compound, starting from the low temperature, the $\chi$ decreases with T and consists two humps at around 30 K ($T_{SR1}$) and 50 K ($T_{SR2}$) which can be attributed to the SR temperatures as reported in the literature [20]. In general such a SR transition in the rare-earth orthoferrites appear due to the transition between the $\Gamma_4 \rightarrow \Gamma_2$ and takes place between the rare-earth ordering temperature ($T_{N1}$) and the $T_{N2}$ (due to antiferromagnetic (AFM) alignment of Fe moments)[12]. It has been believed that the $T_{SR1}$ is due to the competing Zeeman and Van Vleck mechanism [21], which vanishes upon doping the $Fe^{3+}$ site with the $Cr^{3+}$. Above 50 K we do not see any transition until 300 K neither due to an antiferromagnetism nor due to the SR, as though the $Fe^{3+}$ moments couple antiferromagnetically at higher temperatures above 300 K. In order to find the exact value of $T_{N2}$, a high temperature DSC in specific heat mode was used. The inset of Fig. 2(a) shows the DSC graph in a specific heat mode in the temperature range of 400 - 800 K. Essentially, the phase change from an AFM to a paramagnetic (PM) demands an entropy change, which would be reflected as heat flow curve in the DSC. It is evident from the figure that there exists an endothermic peak at 643 K which can be regarded as a Néel temperature ($T_{N2}$). During the phase change, the heat is absorbed as it is evident from the endothermic peak in the DSC.

In contrast, $\chi$ for the $x$ = 0.5 compound consists a cusp at the low temperatures (~ 13 K) which could be due to the ordering of the rare-earth ion, $Ho^{3+}$ ($T_{N1}$). As we go to high temperatures, indeed there exists a transition due to the SR around 150 K, such SR transitions are apparent in the parent compound below 50 K. It is very interesting to note that in the parent compound two such SR transitions are evident below 50 K, while only one prevails in the $x$ = 0.5 compound at 150 K. The shift in the SR transition for the $x$ = 0.5 compound is due to the fact that the SR transition temperature is determined by the complex exchange interaction between the $Fe^{3+}$ and the $Ho^{3+}$ ions [22]. Indeed there would be a competition between the $Ho^{3+}$- $Fe^{3+}$ and the $Fe^{3+}$- $Fe^{3+}$ interactions and if the $Fe^{3+}$- $Fe^{3+}$ interaction is strong, the probability of having the SR transition is at the low temperature. However, if the $Fe^{3+}$- $Fe^{3+}$ is weak the SR transition would takes place at high temperatures. In the present work as we substitute the $Cr^{3+}$ to the $Fe^{3+}$ site, the smaller value of the effective magnetic moment ($\mu_{eff}$) pertinent to the $Cr^{3+}$ ($3.872\mu_B$) compared to that of the $Fe^{3+}$ ($5.916\mu_B$) reduces the contribution of the $Fe^{3+}$ in the $Fe^{3+}$ - $Fe^{3+}$ interaction in the $x$ = 0.5 compound. This indeed leads to the weakening of the $Fe^{3+}$- $Fe^{3+}$ interaction and results in the domination of the $Ho^{3+}$ - $Fe^{3+}$ interaction, which demands a shift in the SR transition to a higher temperature (~ 150 K). Above 150 K, it is apparent that there exists a small peak (clear from 1/χ *vs.* T) at 273 K pertinent to the AFM transition. This also gives an indication that the $Cr^{3+}$ dilutes the $Fe^{3+}$- $Fe^{3+}$ AFM interaction and hence the $T_{N2}$ drastically reduces from 643 K to 273 K for $x$ = 0.5 compound. The observed sudden increase in the susceptibility at low temperatures can be ascribed to the ordering of the paramagnetic rare-earth $Ho^{3+}$ ions.

Fig. 2(b) shows the temperature dependence of the inverse susceptibility (1/χ *vs.* T) obtained for the $x$ = 0.5 compound. After fitting the linear region above $T_{N2}$ (273 K), from the slope $\mu_{eff}$ is obtained and it is found to be $11.619\mu_B$. The estimated value of the $\mu_{eff}$ is found to be close to the theoretical value ($\mu_{theory} = 11.726\mu_B$) calculated from the free ion values of $Ho^{3+}$ ($10.607\mu_B$), $Fe^{3+}$ ($5.916\mu_B$), and $Cr^{3+}$ ($3.872\mu_B$) (spin only values for $Fe^{3+}$ and $Cr^{3+}$) using the equation [23] $\mu_{theory} = \left[\mu^2_{Ho^{3+}} + (1-x)\mu^2_{Fe^{3+}} + x\mu^2_{Cr^{3+}}\right]^{1/2}$ and by assuming their randomness in the paramagnetic phase. The equation $\mu_{Fe^{3+}/Cr^{3+}} = 2\left[S(S+1)\right]^{1/2}$ is used to calculate $\mu_{Fe^{3+}}/\mu_{Cr^{3+}}$ moments and the formula $\mu_{Ho^{3+}} = |g_J|\left[J(J+1)\right]^{1/2}$ is used to calculate $\mu_{Ho^{3+}}$ moments. In the above equations $S$ is the spin sate of $\mu_{Fe^{3+}}/\mu_{Cr^{3+}}$ and $J$ represents the total angular momentum and $g_J$ represents the Landè $g$-factor respectively. For the $x$ = 0.5 compound, the paramagnetic Curie temperature is estimated by extrapolating the high temperature linear region of 1/χ *vs*. T graph to x – axis and it is found to be - 10 K, which indicates an antiferromagnetic behaviour.

The isothermal magnetization loops are measured at 300 K and 5 K in order to get more insights about the magnetic characteristics of $HoFe_{1-x}Cr_xO_3$ ($x$ = 0 and 0.5) compounds. Figure 3(a) and 3(b) shows the magnetic field dependence of the magnetization (M *vs.* H) for $HoFe_{1-x}Cr_xO_3$ ($x$ = 0 and 0.5) compounds measured at 300 K and 5 K respectively. Indeed there exists a loop opening (coercivity $H_C$ = 0.0720 T & remanence $M_r$ = 0.021 $\mu_B$/f.u.) at the origin for $x$ = 0 as it is evident from the inset of Fig. 3(a). In contrast, M *vs.* H variation is linear for $x$ = 0.5 compound, which is in accordance with our earlier observation that this compound consists $T_{N2}$ at around 273 K, above which it is a paramagnet. The loop opening for the $x$ = 0 compound can be ascribed to the weak ferromagnetic behaviour which may be due to the presence of the canted $Fe^{3+}$ moments. Fig. 3(b) shows the M *vs.* H loops at 5 K. It is apparent from the figure that the magnetization consists a loop opening for both the $x$ = 0 and $x$ = 0.5 compounds with a $H_C$ value of 0.2266 T and 0.0743 T respectively. This indeed indicates that with the $Cr^{3+}$ addition, compound turns magnetically soft. The magnetization behaviour at 5 K can be ascribed to the weak ferromagnetic nature of the compounds due to the canted moments in AFM configuration as a result of the Dzyaloshinsky-Moriya antisymmetric exchange interaction [9, 24]. The M *vs.* H can be represented as $\mathrm{M} = \chi_{\mathrm{AFM}}\mathrm{H} + \sigma_s$, where $\sigma_s$ is the saturation magnetization of the weak ferromagnetism, and $\chi_{\mathrm{AFM}}\mathrm{H}$ is the antiferromagnetic contribution [25]. The decrease in magnetization by incorporating $Cr^{3+}$ ions at iron site may be due to the smaller effective moment of the $Cr^{3+}$ ( $3.872\mu_B$ ) in comparison with the $Fe^{3+}$ ($5.916\mu_B$ ).

MC measurements were carried out on a polycrystalline $HoFeO_3$ compound to get the information about the microstructural changes in the material for an applied magnetic field. Fig.4 (a) indicates the room temperature variation of capacitance with respect to the H at a frequency of 100 Hz. It is evident from the figure that the capacitance increases with magnetic field until 0. 5 T and shows diminishing trend until 0.4 T. Further decrease in magnetic field leads to an increase in capacitance until - 0.8 T and continues to decrease until – 0.4 T, below which again it increases until 0 T. Arrows on the graph indicates the sequence for the capacitance behaviour. Indeed there exists a butterfly kind of loop, which resembles the variation of magnetostriction with respect to the magnetic field. Probable reason for the observed MC in the present compound could be due to (a) magnetostriction (b) magnetoelectric coupling [18]. In order to check we have done both the magnetostriction vs. magnetic field measurements and polarization (P) vs. electric field (E) measurements at 300 K. Magnetostriction measurements were carried out using strain gauge method and the present compound doesn't show any magnetostriction for and applied magnetic field as high as 1 T, hinting that the contribution from magnetostriction to MC can be ruled out. Fig. 4(b) shows unsaturated P *vs.* E loops on $HoFeO_3$ at 300 K with a frequency of 100 Hz and at different applied electric fields in the range of 2 – 5 kV/cm, suggests the lossy dielectric behaviour [26]. The polarization of 0.38 $\mu C/cm^2$ is evident at 5 kV/cm. Inset of Fig. 4(b) shows the small opening of P *vs.* E loop measured in the electric field range of 5 kV/cm. Indeed such an observation is puzzling due to the fact that one cannot expect electric polarization for a compound with centrosymmetric space group Pbnm. It has been observed that the spontaneous or the magnetic field induced electric polarization in orthoferrites appears below the antiferromagnetic Néel temperature of the rare earth ion [27]. Concurrently, it has been believed that the coexistence of $Fe^{2+}/Fe^{3+}$ in rare earth iron garnets leads to the movement of charge carriers and local dipoles which may in turn strongly affect their physical properties [28, 29]. Earlier ferroelectric behaviour and magnetocapacitance behaviour has been observed in $Ho_3Fe_5O_{12}$ due to $Fe^{2+}$ [18]. From our XRD it is evident that there exists a peak which corresponds to $Ho_3Fe_5O_{12}$ phase, hence, we believe that a ferroelectric behaviour and MC behaviour that we observed could be due to extra phase that is evident during our phase purity investigation.

**Conclusions**:

In summary, we have successfully demonstrated an increase in spin reorientation transition temperature for the compound with a composition $x$ = 0.5 and the magnetocapacitance behaviour in $HoFeO_3$ compound. The shift in SR transition to high temperature for $x$ = 0.5 compound ascribed to the domination of $Ho^{3+}$ - $Fe^{3+}$ interaction over the $Fe^{3+}$ - $Fe^{3+}$ interaction. On the other hand, smaller effective magnetic moment of $Cr^{3+}$ compared with $Fe^{3+}$ diminishes the Néel temperature as well as the magnetization. In addition to the above, the reduction in the $H_C$ up on the $Cr^{3+}$ addition suggests the magnetically softening of the $HoFeO_3$ compound. Observed magnetocapacitance and ferroelectric behaviour in $HoFeO_3$ is believed due to the lossy dielectric mechanism as a result of an extra phase $Ho_3Fe_5O_{12}$. We believe that the present results would be helpful not only to understand the complex magnetic interactions in orthoferrites but also to develop devices based on them.

**Acknowledgements**

We would like to acknowledge Indian Institute of Technology, Hyderabad and Department of Science and Technology (DST) **(**Project #SR/FTP/PS-190/2012**)** for the financial support.

**References**:

[1] Hui Shen, Zhenxiang Cheng, Fang Hong, Jiayue Xu, Shujuan Yuan, Shixun Cao, and Xiaolin Wang, Magnetic field induced discontinuous spin reorientation in $ErFeO_3$ single crystal, *Appl. Phys. Lett.* **103** (2013) pp. 192404.

[2] K. P. Belov, A. K. Zvezdin, A. M. Kadomtseva, and R. Z. Levitin, Spin-reorientation transitions in rare-earth magnets, *Sov. Phys. Uspekhi* **19** (1976) pp. 574-596.

[3] L. T. Tsymbal, N. K. Dan'shin, V. D. Buchel'nikov, V. G. Shavrov, Magnetoacoustics of rare earth orthoferrites, *Phys. Usp.* **39** (1996) pp. 547 - 572.

[4] A. V. Kimel, A. Kirilyuk, P. A. Usachev, R. V. Pisarev, A. M. Balbashov, and Th. Rasing, Ultrafast non-thermal control of magnetization by instantaneous photomagnetic pulses, *Nature* **435** (2005) pp. 655-657.

[5] R. L. White, Review of Recent Work on the Magnetic and Spectroscopic Properties of the Rare-Earth Orthoferrites, *J. Appl. Phys.* **40** (1969) pp. 1061-1069.

[6] M. Gateshki, L. Suescun, S. Kolesnik, J. Mais, K. Świerczek , S. Short, B. Dabrowski, Structural, magnetic and electronic properties of $LaNi_{0.5}Fe_{0.5}O_3$ in the temperature range 5–1000 K, *Journal of Solid State Chemistry* **181** (2008) pp. 1833-1839.

[7] N. Singh, J. Y. Rhee, S. Auluck, Electronic and Magneto-Optical Properties of Rare-Earth Orthoferrites $RFeO_3$ (R = Y, Sm, Eu, Gd and Lu), *J. Korean Phys. Soc.* **53** (2008) pp. 806-811.

[8] S. Acharya, J. Mondal, S. Ghosh, S.K. Roy, P.K. Chakrabarti, Multiferroic behaviour of lanthum orthoferrites ($LaFeO_3$), *Materials Letters* **64** (2010) pp. 415-418.

[9] S. Iida, K. Ohbayashi, S. Kagoshima, Magnetism of rare-earth orthoferrites as revealed from critical phenomena observation, *J. Phys. Colloques* **32** (1971) pp. C1-654–C1-656.

[10] A. V. Kimel, B. A. Ivanov, R. V. Pisarev, P. A. Usachev, A. Kirilyuk and Th. Rasing, Inertia-driven spin switching in antiferromagnets, *Nature* (*London*) **5** (2009) pp. 727–731.

[11] Zhiqiang Zhou, Li Guo, Haixia Yang, Qiang Liu, Feng Ye, Hydrothermal synthesis and magnetic properties of multiferroic rare-earth orthoferrites, *Journal of Alloys and Compounds* **583** (2014) pp. 21–31.

[12] John C. Walling and Robert L. White, Optical-absorption Zeeman Spectroscopy of $HoFeO_3$, *Phys. Rev. B* **10** (1974) pp. 4737- 4747.

[13] P.D. Battle, T.C. Gibb, A.J. Herod, S.H. Kim, P.H. Munns, Investigation of magnetic frustration in $A_2FeMnO_6$ (A = Ca, Sr, Ba; M = Nb, Ta, Sb) by magnetometry and Mössbauer spectroscopy, *J. Mater. Chem.* **5** (1995) pp. 865-870.

[14] John B. Goodenough, Theory of the Role of Covalence in the Perovskite-Type Manganites [La, *M* (II)] $MnO_3$, *Phys. Rev.* **100** (1955) pp. 564-573, J Kanamori, Superexchange interaction and symmetry properties of electron orbitals, *J. Phys. Chem. solids* **10**(2) (1959) pp. 87-98.

[15] B. Rajeswaran, P. Mandal, Rana Saha, E. Suard, A. Sundaresan, and C. N. R. Rao, Ferroelectricity Induced by Cations of Nonequivalent Spins Disordered in the Weakly Ferromagnetic Perovskites, $YCr_{1-x}M_xO_3$ (M = Fe or Mn), *Chem. Mater.* **24** (2012) pp. 3591-3595.

[16] Vidhya G. Nair, L. Pal, V. Subramanian, and P. N. Santhosh, Structural, magnetic, and magnetodielectric studies of metamagnetic $DyFe_{0.5}Cr_{0.5}O_3$, *J. Appl. Phys.* **115** (2014) pp. 17D728.

[17] N. Kimizuka, A. Yamamoto, H. Ohashi, T. Sugihara, T. Sekine, The stability of the phases in the $Ln_2O_3$-FeO-$Fe_2O_3$ systems which are stable at elevated temperatures (Ln: Lanthanide elements and Y, *J. Solid State Chem.* 48 (1983) pp. 65-76.

[18] Jie Su, Yunfeng Guo, Junting Zhang, Hui Sun, Ju He, Xiaomei Lu, Chaojing Lu, and Jinsong Zhu, Ferroelectric behaviour and magnetocapacitance effect caused by $Fe^{2+}$ in $Ho_3Fe_5O_{12}$, *Ferroelectrics* 448 (2013) pp. 71-76

[19] Ganesh Kotnana and S. Narayana Jammalamadaka, Band gap tuning and orbital mediated electron-phonon coupling in $HoFe_{1-x}Cr_xO_3$ $(0 \leq x \leq 1)$, *J. Appl. Phys.* **118** (2015) pp. 124101.

[20] Shujuan Yuan, Ya Yang, Yiming Cao, AnhuaWu, Bo Lu, Shixun Cao, Jincang Zhang, Tailoring complex magnetic phase transition in $HoFeO_3$, *Solid State Communications* **188** (2014) pp. 19–22.

[21] O. Nikolov, T. Ruskov, G.P. Vorobyov, A.M. Kadomtseva, I.B. Krynetskii, A new mechanism of the spin reorientations in $HoFeO_3$, *Hyperfine Interact.* **54** (1990) pp. 623-626.

[22] S. Venugopalan, M. Dutta, A. K. Ramdas, and J. P. Remeika, Magnetic and vibrational excitations in rare-earth orthoferrites: A Raman scattering study, *Phys. Rev. B* **31** (1985) pp. 1490-1497.

[23] A. J. Dekker, *Solid State Physics*, Prentice-Hall, New York (1970).

[24] I. Dzyaloshinsky, A thermodynamic theory of “weak” ferromagnetism of antiferromagnetics, *J. Phys. Chem. Solids* **4** (1958) pp. 241-255.

[25] A. Dahmani, M. Taibi, M. Nogues, J. Aride, E. Loudghiri, and A. Belayachi, Magnetic properties of the perovskite compounds $YFe_{1-x}Cr_xO_3$ $(0.5 \leq x \leq 1)$, *Mater. Chem. Phys.* **77** (2003) pp. 912-917.

[26] J. F. Scott, Ferroelectrics go bananas, J. Phys.: Condens. Matter 20 (2008) 021001.

[27] Y. Tokunaga, S. Iguchi, T. Arima, and Y. Tokura, Magnetic-Field-Induced Ferroelectric State in $DyFeO_3$, *Phys. Rev. Lett.*101 (2008) pp. 097205.

[28] J. Su, X.M. Lu, C. Zhang, J. T. Zhang, H. Sun, C. Zhang, Z. H. Jiang, C.C. Ju, Z. J. Wang, F. Z. Huang, and J. S. Zhu, The effect of $Fe^{2+}$ ions on dielectric and magnetic properties of $Yb_3Fe_5O_{12}$ ceramics, *J. Appl. Phys*. 111 (2012) pp. 014112.

[29] M. A. Subramanian, Tao He, Jiazhong Chen, Nyrissa S. Rogado, Thomas G. Calvarese, and Arthur W. Sleight, Giant room temperature magnetodielectric response in the electronic ferroelectric $LuFe_2O_4$, *Adv. Mater*. 18 (2006) pp. 1737-1739.

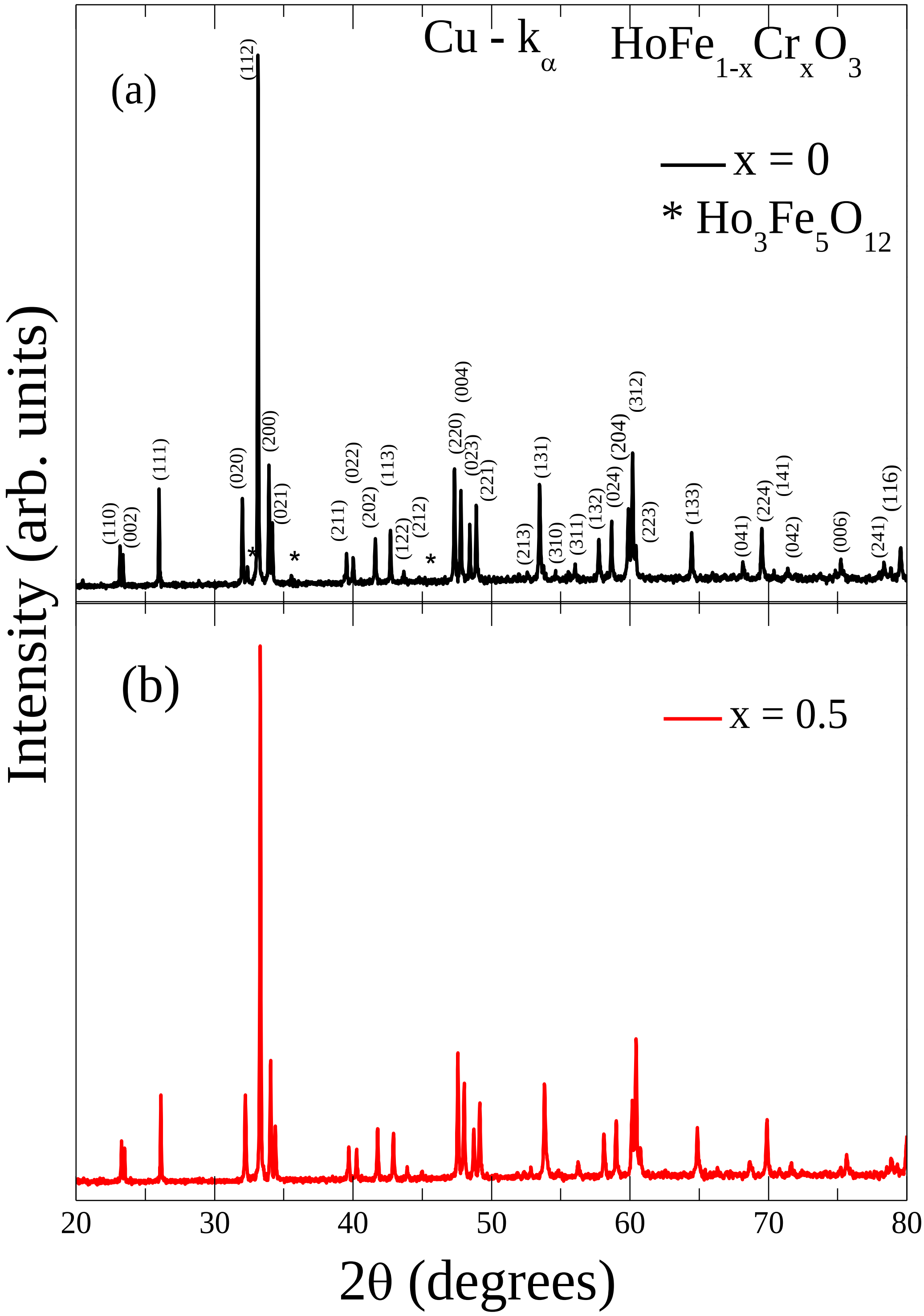

**Fig. 1:** Room temperature powder XRD patterns of $HoFe_{1-x}Cr_xO_3$ (a) $x = 0$ (b) $x = 0.5$.

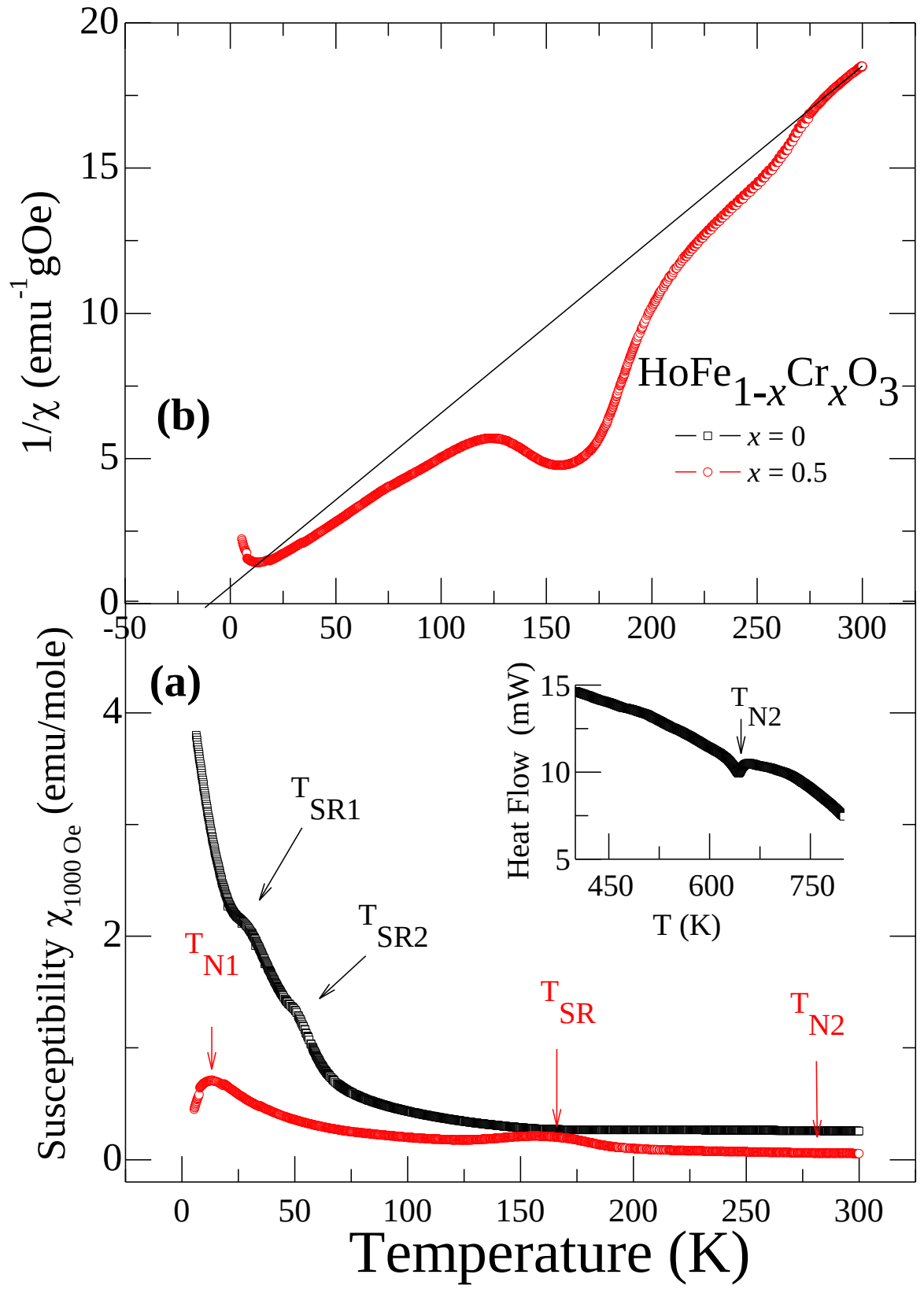

**Fig. 2:** (a) Susceptibility variation for the compounds $HoFe_{1-x}Cr_xO_3$ ($x$ = 0, 0.5) in the temperature range 5 – 300 K. Inset of (a) shows DSC heat flow curve for $HoFeO_3$ in the temperature range of 400-800 K (b) Inverse susceptibility *vs.* T for the compound $HoFe_{0.5}Cr_{0.5}O_3$. It is evident that above $T_{N2}$ (273 K), graph shows linear behaviour.

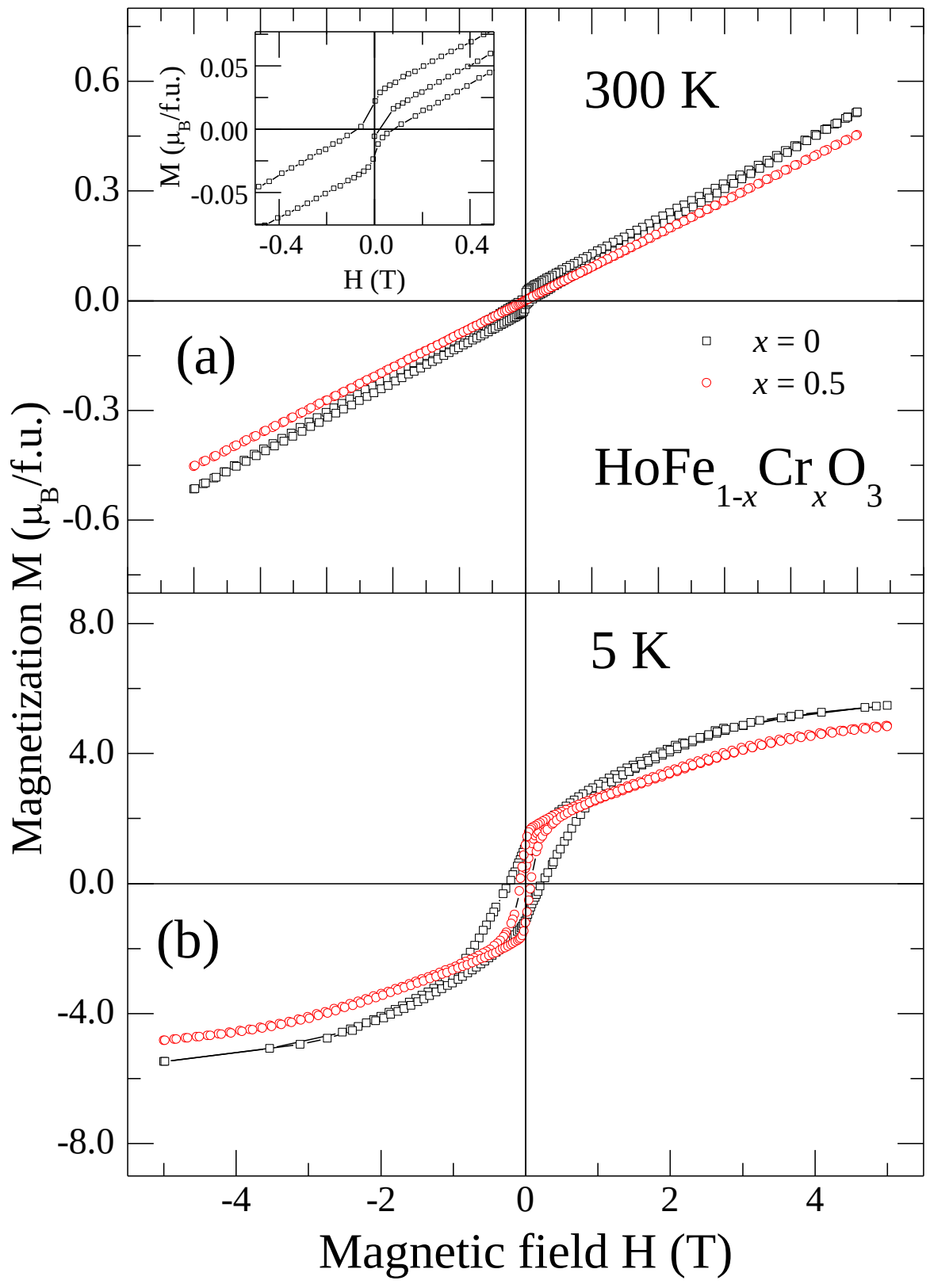

**Fig. 3:** Magnetic field *vs*. Magnetization graph for $HoFe_{1-x}Cr_xO_3$ ($x$ = 0, 0.5) compounds (a) at 300 K. Inset shows the expanded view for $x$ = 0 compound (b) Magnetization graphs at 5 K for $x$ = 0 and $x$ = 0.5 compounds respectively.

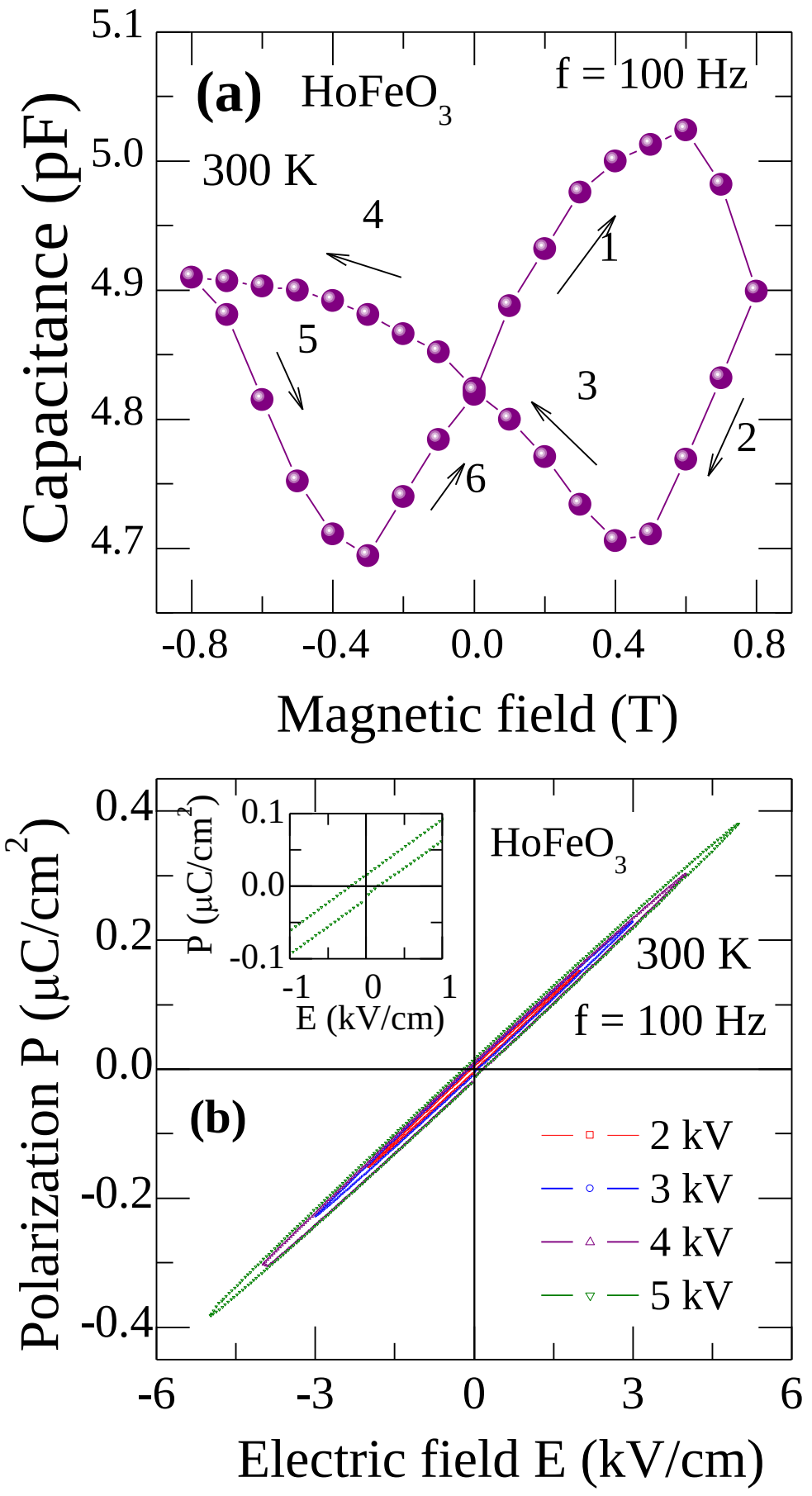

**Fig 4:** (a) Magnetic field *vs.* capacitance of $HoFeO_3$ at 300 K with 100 Hz frequency (b) Electric polarization (P) *vs.* electric field (E) hysteresis loops of $HoFeO_3$ measured at 300 K with 100 Hz frequency.